\documentstyle[multicol,aps]{revtex}

\begin{document}

\title{Ellipsoidal deformation of vertical quantum dots} 
\author{D.G. Austing\footnote{Email:austing@will.brl.ntt.co.jp}, S. Sasaki, S. Tarucha\footnote{present adress: University of Tokyo, Physics 
Department, 7-3-1 Hongo, Bunkyo-ku, Tokyo 113-0033 Japan}}

\address{NTT Basic Research Laboratories,
3-1 Morinosato Wakamiya, Atsugi,
 Kanagawa 243-0198, Japan}
\author{S.M. Reimann\footnote{Email: reimann@phys.jyu.fi}, M. Koskinen and M. Manninen}
\address{University of Jyv\"askyl\"a, Physics Department,
PO Box 35, 40351  Jyv\"askyl\"a, Finland}
\maketitle
\begin{abstract}
Addition energy spectra at 0 T of circular and ellipsoidally deformed 
few-electron vertical quantum dots are measured and compared to results of 
model calculations within spin-density functional theory. Because of the 
rotational symmetry of the lateral harmonic confining potential, circular 
dots show a pronounced shell structure. With the lifting of the single-
particle level degeneracies, even a small deformation is found to radically 
alter the shell structure leading to significant modifications in the addition 
energy spectra. Breaking the circular symmetry with deformation also 
induces changes in the total spin. This ``piezo-magnetic'' behavior 
of quantum dots is discussed, and the addition energies for a set of 
realistic deformation parameters are provided. For the case of the 
four-electron ground state at 0 T, a spin-triplet to spin-singlet transition 
is predicted, i.e. Hund's first rule no longer applies. Application of a 
magnetic field parallel to the current confirms that this is the case, and 
also suggests that the anisotropy of an elliptical dot, in practice, may be 
higher than that suggested by the geometry of the device mesa in which the dot 
is located. 
\end{abstract}

\begin{multicols}{2}
\narrowtext

\newpage

\section{Introduction}

In atomic physics, the mean-field method describing the 
motion of electrons confined in the three-dimensional spherically symmetric
Coulomb potential of the nucleus provides an impressively powerful tool to 
explain the chemical inertness and special stability of the noble gases.
The well-known atomic shell structure is a consequence of the fact 
that the atomic levels 1s, 2s, 2p, 3s, 3p, ...  show a ``bunchiness'' in their 
distribution as a function of energy. Particular stability of the electronic 
system is reached when a bunch of such levels is fully occupied. If then one 
more electron is added, the electron configuration would involve a singularly 
occupied orbital from the next higher shell, and consequently, the system is 
then less stable. Shell filling is thus reflected by large maxima in the 
ionization energy for atomic numbers 2, 10, 18,~..., corresponding to the 
nobel gas atoms He, Ne, Ar, ...~. In the mid-shell regions, large level 
degeneracies occur as a consequence of the spherical symmetry of the 
confining potential of the atomic nucleus. The mid-shell levels are then 
filled according to Hund's rules, in particular maximizing the total electron 
spin for half-filled orbitals~\cite{weissbluth}.

A shell structure is not only unique to atoms, but actually is a
recurring property in finite fermion systems with high symmetry~\cite{bm}.
It equally explains the occurence of ``magic'' proton and 
neutron numbers in the binding energies of nuclei, and more recently the 
discovery of ``magic'' atom numbers in metal clusters~\cite{clusters} -- small 
aggregates of metal atoms in which delocalized valence electrons move in the 
positive charge background of the ions. Fundamentally, in contrast to atoms, 
however, both mid-shell nuclei and clusters deform their mean field rather 
than obey Hund's rules.

	The two-dimensional analogue to the atom with its static $1/r$ 
radial Coulomb confinement due to the nucleus can  be realized in small 
semiconductor devices: artificial semiconductor atoms based on quantum
dot technology. Clean, well-defined, and highly symmetric vertical quantum 
dots (``islands'') can now be made so small that the dot size is comparable to 
the Fermi-wavelength~\cite{AUSTING,AUSTING2,TARUCHA2}. Typical micrographs of 
micron-sized device mesas incorporating these dots are shown in 
Fig.~\ref{fig:1}. The lateral electrostatic confinement originates from 
side-wall depletion, and this (and the effective dot size) can be controlled 
or ``squeezed'' by the action of a Schottky gate wrapped around the mesa in 
the vicinity of the dot to the degree that the number of electrons trapped 
on the dot can be changed one-by-one. Also the few-electron regime is readily 
accessible, and then the resulting confinement is well approximated by a 
parabolic $r^2$ potential. Electron phenomena in related semiconductor 
quantum dot structures continue to attract much attention
~\cite{KASTNER,MEIRAV,ASHOORI}. As they exhibit atomic-like properties, 
such as a shell structure and shell filling in accordance with Hund's first 
rule, the vertical quantum dots, whose heterostructure barriers are both 
abrupt and thin, can be regarded as artificial atoms whose ground and 
excited states can be probed electrically by single electron tunneling 
spectroscopy in order to perform novel ``atomic physics'' experiments in the 
few-electron regime~\cite{TARUCHA,KOUWENHOVEN}. 

	When an arbitrarily small bias, $V$, is applied across the dot between 
the metal contact on top of the device mesa and the substrate contact (these
are often refered to as the source and drain contacts), the ground states of 
an $N-$electron quantum dot weakly coupled to the contacts can be investigated
directly by monitoring the current flowing vertically through the dot at or 
below 0.3 K as the voltage on a single gate, $V_g$, surrounding the dot is 
varied. When no current flows (Coulomb blockade), $N$ is well defined. On the 
other hand, when current flows the number of electrons can oscillate between 
$N$ and $N+1$. With the gate, $N$ can be increased one-by-one starting from 
zero by making $V_g$ more positive, so a series of sharp current peaks due 
to the charging of the dot (Coulomb oscillations) can be observed. For a 
large dot containing many electrons, the Coulomb oscillations are usually 
periodic because the single electron charging energy is determined classically
just by the total dot capacitance. For a dot containing just a few electrons 
both quantum effects reflecting the underlying symmetry of the confining 
potential, and the details of the electron-electron interactions become 
important as the dot size is reduced. This leads to modifications of the 
Coulomb oscillations, so they are no longer expected to be periodic
~\cite{TARUCHA2,TARUCHA}.

	To date, we have mainly focused on the properties of dots in circular 
mesas which have diameters of typically 0.4 to 0.7 microns. For a magnetic 
field parallel to the current, the measured ground states between 0 T and
about 4 T for $N<20$ in these disk-shaped dot can be well accounted for by a 
single-particle picture based on the Darwin-Fock spectrum for a circular
two-dimensional harmonic confining potential, a constant interaction, and 
corrections at 0 T due to exchange, i.e. Hund's first rule
~\cite{TARUCHA2,TARUCHA}. At higher fields beyond about 4 T, the evolution 
of ground states (and also the excited states) for $N<6$ can be understood 
in terms of many-body effects~\cite{KOUWENHOVEN}.

	The main theme of this article concerns the effect of geometrically 
distorting a circular dot into an elliptical (anisotropic) dot. Previously, 
we have briefly reported some properties of elliptical dots
~\cite{TARUCHA2,SASAKI}. Here, we present a more detailed study of the 
addition energies and include their magnetic-field dependencies. The 
experimental data are compared to model calculations. We survey general 
trends, and examine basic assumptions about the nature of the deformed dots. 

	A perfectly circular dot possesses full rotational symmetry. This high 
symmetry leads to maximal level degeneracy of the single-particle 
two-dimensional states for parabolic confinement, and this emphasises atomic-like 
properties~\cite{TARUCHA2}. This level degeneracy at 0 T for a circular dot 
is evident in the single-particle spectrum in Fig.~\ref{fig:2}(a), and 
consecutive filling of each set of degenerate states is directly responsible 
for the characteristic shell structure with ``magic'' numbers 
$N= $2, 6, 12, 20,~... . Furthermore, Hund's first rule accounts 
for the parallel filling of electrons amongst half-filled degenerate 
states in a shell at numbers $N= $4, 9, 16,~... due to an exchange effect. 
Breaking the circular symmetry by deforming the lateral confining potential 
lifts the degeneracies of the single-particle levels present in a disk-shaped 
dot. This destroys the shell structure for a circle, and modifies other 
atomic-like properties~\cite{MADHAV}.

	The sequence of spectra in Fig.~\ref{fig:2} also introduces two key 
points in our subsequent arguments. Firstly, as the deformation is 
gradually increased, (a) to (d), degeneracies of the single-particle states
at 0 T are generally removed. Nevertheless, accidental degeneracies 
can occur at certain ``magic'' deformations, e.g. (b) and (c), leading to 
subshell closures, provided the confining potential is still perfectly 
parabolic. The resulting patterns, however, are very different from 
that for the circular case, (a), and in practice may be hard to observe. 
Secondly, a weak magnetic field parallel to the current can also induce 
level degeneracies in both circular and elliptical dots when single-particle 
levels cross at finite field, but here too, any shell structure at a 
particular field is of a lower order and less apparent than that for the 
circle at 0 T~\cite{MADHAV}.

	While illustrative, ultimately any modelling of the behavior of real 
dots must go beyond a system of $N$ non-interacting electrons confined by a
two-dimensional harmonic oscillator, i.e.~a single-particle picture, as 
employed to generate the spectra in Fig.~\ref{fig:2}~\cite{MADHAV}, and 
include Coulomb interactions which can lift certain degeneracies at 0 T. 
Numerical diagonalization of the full Hamiltonian matrix has recently been 
successfully employed to calculate basic electronic properties of dots with 
anisotropic confining potentials~\cite{EZAKI,EZAKI2}. Such ``exact'' numerical
calculations, however, are limited to only a few confined particles. In order 
to study dots confining a larger number of electrons 
we apply spin-density functional theory at 0 T. This powerful technique, 
which explicitly incorporates the electron-spin interactions, has lead to a 
number of interesting predictions for the ground state structure of quantum 
dots, although there is a continuing discussion as to the interpretation
of so-called spin-density 
waves (SDW)~\cite{LEE,koskinen,steffens,serra,reimann1,Hirose}. Both ``exact'' 
numerical calculations and spin-density functional theory predict subtle 
changes in the addition energy spectra, and transitions in the spin-states 
as deformation is varied -- even for a weak deformation. An example of the 
latter is the breakdown of the conditions for which Hund's first rule 
applies for four electrons, and this marks a transition from a spin-triplet to 
a spin-singlet configuration, i.e. states are consecutively filled by spin-up 
and spin-down electrons.  

\section{Experimental setup}

	The vertical quantum dots under focus in the following are fabricated 
by electron-beam lithography, and a two step etching technique to make 
circular or rectangular sub-micron mesas from one special
GaAs/Al$_{0.22}$Ga$_{0.78}$As/\-In$_{0.05}$Ga$_{0.95}$As/\-Al$_{0.22}$
Ga$_{0.78}$As/\-GaAs double barrier heterostructure (DBH). Full details of 
the device fabrication, and the material parameters are given elsewhere
~\cite{AUSTING,AUSTING2,TARUCHA2,TARUCHA}. A single Schottky gate is placed 
around the side of the mesa close to the DBH. We discuss one circular mesa 
with a nominal top contact diameter, $D$, of 0.5 $\mu$m (W), and three 
rectangular mesas with a top contact area $(L\times S)$ 0.55 $\times  0.4$  
$\mu$m$^2$ (X), 0.65 $\times  0.45$  $\mu$m$^2$ (Y), and 0.6 $\times  0.4$  
$\mu$m$^2$ (Z). $L(S)$ is the nominal dimension of longest (shortest) side 
of the top contact. Fig.~\ref{fig:1} shows typical scanning electron 
micrographs of a circular mesa, and a rectangular mesa taken immediately 
after the depositon of the Schottky gate metal surrounding the mesa. For 
the rectangular mesas, an intuitively simple way to classify them is to 
define a geometric parameter, $\beta $, to be the ratio $L/S$. For X, Y, and 
Z respectively $\beta $ is nominally 1.375, 1.44 and 1.5. Due to a slight 
isotropic undercut resulting from the light wet etch during the formation of 
the mesa~\cite{AUSTING,AUSTING2}, the area of the mesas, as revealed by the 
micrographs, is a little less than that of the top contact, so realistic 
values for $\beta $ are estimated to be about 5\% larger than the values 
quoted.

	Fig.~\ref{fig:1} also schematically shows the slabs of semiconductor 
between the two Al$_{0.22}$Ga$_{0.78}$As tunneling barriers, and the resulting
dots bounded by the shaded depletion region for the circular and rectangular 
mesas. The thickness of the In$_{0.05}$Ga$_{0.95}$As slab is determined by 
the separation between the well defined heterostructure tunneling barriers 
(approximtely 100{\AA}). The slab is sufficiently thin that all electrons 
are in the lowest state in the vertical direction parallel to the current. 
The lateral confining potential due to the side wall depletion further 
restricts electrons to the center of the slab, thus defining the dot region.
We note that in our devices, the extent of the lateral depletion region in the
vicinity of the dot is largely determined by the electron density in the 
n-doped GaAs regions above and below.  

	The lateral harmonic confining potential of the dot in the circular 
mesa has circular symmetry of a sufficiently high degree that degenerate sets 
of states can systematically form in the disk-shaped dot
~\cite{TARUCHA2}. These states can be labelled by the quantum numbers 
($n$,$l$), where $n$ is the radial quantum number (=0, 1, 2, ...), and $l$ is 
the angular momentum quantum number $(=0,\pm 1,\pm 2, ...)$. Each state can 
hold a spin-up electron and a spin-down electron. At 0 T the $2n+|l|+1$-th 
shell is made up of $2n+|l|+1$ degenerate single-particle states. Each 
degenerate set of states can be regarded as a shell of an artificial atom, 
and this is the origin of the 2, 6, 12, 20, ... ``magic'' numbers. The first 
shell consists of the (0,0) level, the second shell of the (0,1) and (0,-1) 
levels, the third shell of the (0,2), (1,0) and (0,-2) levels, and so on. 
For the circular dots we typically study, the lateral electrostatic 
confinement energy seperating these degenerate sets of single-particle 
states, $E_Q$, can be as large as 5 meV in the few-electron 
limit~\cite{KOUWENHOVEN}. Neglecting an arbitary constant, the energy of
single-particle state ($n$,$l$) is $(2n+|l|+1)E_Q$. The effective lateral 
diameter can be ``squeezed'' from a few thousand Angstroms for $N$ of 
approximately 100 down to 0 {\AA} for $N=0$ by making the gate voltage more 
negative~\cite{TARUCHA2,TARUCHA,KOUWENHOVEN}. We stress that crucially the 
``squeezing'' action of the gate, and indeed application of a magnetic field 
parallel to the current, preserves the circular symmetry of a disk-shaped
dot. Consequently, atomic-like properties should be particularly robust and 
evident in circular dots.  

	For a rectangular mesa, the lateral confining potential of the dot is 
expected to be elliptical-like due to rounding at the corners provided the 
number of electrons in the dot is not too large (in which case it may be more 
rectangular-like with rounded corners), or too small. Right at ``pinch-off'', 
$(N\rightarrow 0)$, it may even become more circular-like, i.e. the 
elliptical-shape of the confining potential may be changing in a complex way
~\cite{TARUCHA2,SASAKI}. Assuming the confining potential is perfectly
parabolic, we can choose to characterize the ``ellipticity'' 
by a deformation parameter, $\delta =E_S/E_L$. Here, $E_S (E_L)$ is 
the confinement energy at 0 T along the minor (major) axis ($E_S>E_L$). The 
states in the elliptical dot are now labelled by the quantum 
numbers ($n_L$,$n_S$), where $n_L$ ($n_S$) is a quantum number (=0, 1, 2, ...)
associated with the energy parabola along the major (minor) axis
~\cite{MADHAV}. Again neglecting an arbitary constant, the energy of 
single-particle state ($n_L$,$n_S$) is ($n_L$+1/2)$E_L$+($n_S$+1/2)$E_S$.

	For a perfectly circular mesa, we can trivially generalize our 
definition of the deformation parameter so that $\delta =\beta =1$. 
On the other hand, for the rectangular mesas, there is no simple 
correspondance between $\beta $, a ratio of lengths characteristic of the 
top metal contact which is independent of gate voltage (or $N$), and 
$\delta $, a ratio of energies characteristic of the dot in the mesa
which is in fact dependent on the gate voltage (or equivalently $N$), i.e. 
``accidental'' degeneracies at ``magic'' deformations will be hard to see 
over an extended range of $N$, and in any case may be lifted if the 
confinement potential is not completely parabolic. Nevertheless, at this 
stage, we start by assuming that $\beta $ is a measure of $\delta $, and 
thus one might expect $\delta_Z>\delta_Y>\delta_X>\delta_W$. We are not
saying that $\delta =\beta $ for the ellipses, and indeed even for the
simplest possible model of uniform depletion spreading due to the action of
the gate, we would expect $\beta $ to underestimate $\delta $. We furthermore 
assume in the following model calculations, for simplicity, that the 
``squeezing'' action of the gate does not alter $\delta $. We will
examine these assumptions in light of the experimental and theoretical 
data presented. Note that the application of a magnetic field parallel 
to the current effectively reduces $\delta $ as seen by the confined 
electrons in the limit of a very high field, where it approaches 
unity. 

\section{Addition energy spectra for circular and deformed dots}

	In Fig.~\ref{fig:3}, the change (formally the second difference) in 
the electro-chemical potential, $\mu(N+1)-\mu(N)=\Delta _2(N)$, which can also
be regarded as a capacitive energy~\cite{LEE}, is plotted as a function of 
electron number, $N$, up to $N=17$ for (a) W, (b) X, (c) Y, and (d) Z at 0 T. 
The traces are offset vertically by 3~meV for clarity. The $N$th current 
peak position in gate voltage, V$_g$, at a very small bias ($\ll 1$ mV), i.e. 
measured in the linear conductance regime, at or below 0.3 K reflects 
$\mu (N)$, the electro-chemical potential of the ground state for $N$ 
electrons, or equivalently the ``addition energy'' to place an extra electron 
on a dot with $N-1$ electrons.$\Delta _2(N)$ then mirrors directly the 
spacing in gate voltage between the $N+1$th and the $N$th current 
peaks~\cite{TARUCHA}. $\Delta _2(N)$ is actually the half-width of the 
$N$th Coulomb diamond, the diamond-shaped region in the $V-V_g$ plane 
in which current is blocked between the $N$th and the $N+1$th current 
peaks. $\Delta _2$ contains contributions from the single-electron charging 
energy and changes in the single-particle energy, $E_Q$
~\cite{TARUCHA,KOUWENHOVEN}.   

       At 0 T, for the circle W, $\Delta _2(N)$ is strongly 
dependent on $N$, and  a very clear characteristic shell structure 
is evident in Fig.~\ref{fig:3}(a)~\cite{TARUCHA}. 
Particularly large peaks ($N=2,6,12$), and relatively 
large peaks ($N=4,9,16$) are indicated. The result from a local spin density 
approximation (LSDA) calculation discussed below is also included for 
comparison~\cite{reimann1}. $N=$2,6, and 12 are the first three ``magic''
numbers for a circular two-dimensional harmonic potential which mark 
completion of the first three shells (containing respectively 1, 2 and 3 
degenerate zero-dimensional single-particle states or equivalently 2, 4 and 6 
electrons). The peaks at $N=$ 4, 9, 16 arise as a consequence of exchange 
effects which are enhanced at half-full shell filling with same-spin 
electrons for the 2nd, 3rd, and 4th shells respectively
~\cite{TARUCHA}. This shell structure should be clear (and this is generally 
the case in practice for $N<20$) as long as: $i)$ the two-dimensional lateral 
potential remains radially parabolic, and rotationally symmetric to a fairly 
high degree, $ii)$ $E_Q$ is comparable to, or larger than, the Coulomb 
interaction energy, and $iii)$ the effect of screening is not significant. 

	For the circular mesa W, it is also evident that as $N$ is decreased, 
$\Delta _2(N)$ generally becomes larger due to the increase of the Coulomb 
interaction when the dot is ``squeezed''. This observation also holds for the
rectangular mesas, but there are no prominant maxima at $\Delta _2$(2,6,12). 
The shell structure for the disk-shaped dot has now become disrupted or 
``smeared out'', and this can be attributed directly to the lifting of the 
degeneracies of the single-particle states that are present in a circular dot
~\cite{TARUCHA2,TARUCHA,SASAKI}. In other words, deformation kills the shell 
structure for a circle, and even quite a small deformation can make a 
big difference. This is evident from the three traces, (b) to (d) in 
Fig.~\ref{fig:3}, but there are major difficulties in discussing specific 
details. As noted earlier, in practice, right at ``pinch-off'', $\delta $ 
may actually tend towards unity~\cite{TARUCHA2,SASAKI}, but more generally 
$\delta \approx \beta $ may be unreliable. Also, even for two circular dots 
which have a clear shell structure in the few-electron limit, the absolute 
values of $\Delta_2$(N) can vary from dot to dot, i.e. the precise 
details are device dependent, and beyond the third shell only a few devices 
show the expected behavior clearly~\cite{TARUCHA}. Lastly, even if 
$\delta $ could be determined accurately, $\Delta _2$(N) strictly 
speaking can only be fairly compared if the ``areas'' of the dots are 
comparable, as in the classical limit $\Delta _2(N)$ is determined by the 
overall dot capacitance~\cite{TARUCHA2}. Based on the nominal sizes
of the mesas, and in line with the trends of the ``pinch-off'' gate voltage 
as identified by the position of the first current peak, elliptical dots X, 
Y, and Z respectively may have ``areas'' 1.1, 1.5 and 1.2 larger than that 
of the circular dot W. Thus to sensibly discuss details, like the spin-states,
even generally, we first calculate $\Delta_2(N)$ at 0 T for a range of 
$\delta $ values in line with those suggested by the $\beta $ values of the 
mesas X, Y, and Z. We then compare with, and look for patterns in, the 
experimental data at 0 T, before looking at the magnetic field dependence 
for confirmation of trends, and whether $\delta \approx \beta $ is reasonable. 

\section{Mean-field model for circular and elliptical quantum dots}

	We next aim to model the changes due to the deformation of the 
lateral confinement to the shell structure of the quantum dots at 0 T by 
applying the methods of spin-density functional theory (SDFT). We will briefly 
address different aspects of the spin structure relevant to the deformed 
quantum dots.

\subsection{The method}

	To obtain the ground-state energies and densities for $N$ electrons 
confined in an externally imposed potential, we solve the spin-dependent 
single-particle Kohn-Sham (KS) equations~\cite{kohnsham}
\begin{equation}
\left[
-{{\hbar^2}\over{2m^*}}\nabla^2 _{\bf r} +
V_{\rm eff}^\sigma ({\bf r})\right]
\psi_{i,\sigma}({\bf r})
=\epsilon_{i,\sigma}\psi_{i,\sigma}({\bf r})
\label{kseq}
\end{equation}
in a plane-wave basis to avoid any symmetry restrictions. 
In Eq.(\ref{kseq}), the  index $\sigma $ accounts 
for the spin ($\uparrow$ or $\downarrow$), and ${\bf r}=(x,y)$.
The effective mean-field potential,
$V_{\rm eff}^\sigma ({\bf r})$, contains contributions from the 
external harmonic confining potential,
the Hartree potential of the electrons,
and the functional derivative of the local
exchange-correlation energy, for which we use 
the approximation of Tanatar and Ceperley~\cite{tantar}  
(see also~\cite{koskinen,reimann1} for details).
The electrostatic confinement due to the lateral depletion region imposed 
by the side wall and the Schottky gate is approximated by a two-dimensional 
anisotropic harmonic oscillator with 
frequencies $\omega _x=\omega \sqrt{\delta }$ and 
$\omega _y=\omega /\sqrt{\delta }$, 
\begin{equation}
V_{ext}(x,y)={1\over 2}m^*
\omega ^2 \left( \delta x^2+ {1\over \delta }y^2 
\right)~. ~
\end{equation}
The ratio of the oscillator frequencies, 
$\delta = \omega _x/\omega _y$, thus 
defines the ratio of semiaxes of the ellipsoidal equipotentials. We 
impose the constraint, $\omega^2$=$\omega _x\omega _y$, which is equivalent to 
conserving the area of the quantum dot with deformation~\cite{reimann1}. 
The $x$ and $y$-axes are indicated in the schematic diagram for the elliptical 
dot in Fig.~\ref{fig:1}. With this convention, the above defined $E_S$ and 
$E_L$ respectively correspond to ${\hbar}\omega _x$ and ${\hbar}\omega _y$. 
In the model we present the dot is assumed to be well isolated from its 
surroundings, so any effects due to the presence of the gate and the 
neighboring conducting regions are neglected. Likewise, screening and non-
parabolicity effects inside the dot, which become more important for large 
N, are not considered.  

	For $\delta =1$, a circular shape for the quantum dot is obtained, 
whereas $\delta > 1$ corresponds to an ellipsoidally deformed quantum dot. The
strength, $\omega $, of the external parabolic confinement leading to an 
average particle density, $n_0=1/(\pi r_s^2)$, 
in a circular dot is approximated 
by $\omega ^2 = e^2 / (4 \pi \epsilon _0 \epsilon 
m^* r_s^3\sqrt{N})$~\cite{koskinen}. 
Minimizing the energy density functional by self-consistently solving the 
above KS equations, Eq.~(\ref{kseq}), ground state energies, 
$E(N,\delta)$, are obtained for different electron numbers and deformation 
parameters. Full technical details are given elsewhere
~\cite{koskinen,reimann1}, and here we report only the results. We 
emphasise that from recent measurements, it is clear that as $N$ increases 
the confinement weakens in such a way that the particle density tends to a 
constant~\cite{AUSTING4}. This is implicit in our model, as for any given 
value of $r_s$, the oscillator frequency $\omega $, and the related 
frequencies $\omega _x$ and $\omega _y$, decrease with increasing $N$. 
$\delta $ is also kept constant for simplicity, although $\delta $ is 
expected to vary with $N$ in practice. 

	Although strictly speaking the dot is located in 
In$_{0.05}$Ga$_{0.95}$As, we take for values of the effective mass, $m^*$, 
and dielectric constant, $\epsilon $, those for GaAs -- namely 0.067
and 13.1 respectively. There are no fitting parameters in the equations, and 
only a suitable choice for $r_s$ is required to generate the addition energy 
spectra. The value of $r_s=1.5$$a_B^*$ used in the model calculations is 
realistic as the value estimated experimentally for a circular quantum dot 
is 1.3 to 1.4$a_B^*$~\cite{AUSTING4}. 
$a_B^*=\hbar ^2 (4\pi \epsilon _0 \epsilon )/m^* e^2$ is an effective atomic 
unit, which for GaAs is about 103$\AA $. $r_s=1.5a_B^*$ in the model 
presented here corresponds to an effective confinement energy, $E_Q$, for 
$N=1$ of about 5.7~meV. This value is consistent with the upper limit of 
$E_Q$ observed in practice (about 5~meV), and justifies the $E_Q$=3~meV 
value as a reasonable average for calculating the simple single-particle 
spectra shown in Fig.~\ref{fig:2} for the first ten levels. 

	We point out that the SDFT calculations described here, as well as
those performed by Hirose and Wingreen~\cite{Hirose}, are strictly 
two-dimensional, so the strength of the Coulomb interactions may be 
overestimated, i.e. the possibility of charge spreading out in both the 
$x$-$y$--plane, and along the vertical direction parallel to the current to 
minimize the Coulomb energy is neglected. Equivalently, anisotropic extension 
of the electron wavefunctions along the major axis is ignored~\cite{TARUCHA2}. 
In practice, screening by the metal contacts surrounding a dot is also 
believed to reduce the influence of Coulomb interactions. The three-dimensional
model of Lee et al.~\cite{LEE} does incorporate self-consistent solution of 
the Poisson equation into a SDFT calculation, but because they use different 
expressions for the exchange-correlation energy, and considerably higher 
values for $E_Q$, $E_L$, and $E_S$ (up to 20~meV), a direct comparision with 
their results is not easy. Nevertheless, they find that electrons strongly 
confined in the vertical direction have a very strong two-dimensional 
character, and both approaches lead to the same qualitative conclusions. 
Namely, the distinct shell-structure for a circle, as well as the spin-states, 
at 0 T are strongly modified with deformation.      

\subsection{$\Delta _2(N)$ for elliptical dots: Results from LSDA calculations}

	We now make a simple comparison between the experimentally measured 
traces for the change in the electro-chemical potential, $\Delta _2(N)$, 
with those modelled theoretically. Fig.~\ref{fig:4} shows $\Delta _2(N)$, 
derived from the self-consistent ground-state energies, $E(N,\delta )$. The 
energies are obtained by self-consistently solving the KS-equations starting 
from different initial guesses for the effective KS-potential for the spin-up 
and spin-down particles. The initial potentials 
are chosen completely arbitrarily by 
just putting small random numbers on to the lattice points. The calculations 
are started from four such guesses. For two of them, the spin-up and spin-down
initial guesses are shifted in value in order to to search for states with 
non-zero total spin for even-$N$. This is important in order to find the 
ground state amongst all possible spin configurations with a high degree of 
certainty~\cite{koskinen,reimann1}.

	The lowest trace in Fig.~\ref{fig:4} gives $\Delta _2(N)$ for the
circular dot ($\delta $=1, i.e. zero deformation). As expected, the circular-
shaped confinement produces a spectrum with the familiar shell structure 
for a two-dimensional harmonic oscillator, with shell closures at the 
``magic'' numbers $2, 6, 12,$ and 20. At the average particle density 
corresponding to $r_s=1.5a_B^*$, these ``magic'' numbers arise from large 
gaps at the Fermi surface and paired spins in each non-degenerate level, so 
the total spin is zero $(S=0)$. We note that within this mean-field model, 
spin-density wave (SDW) states are not expected for these particular 
spin-zero states~\cite{koskinen,reimann1}.

	In Fig.~\ref{fig:3}(a), for $\delta =1$, the experimental and 
theoretical traces can be directly compared. The agreement is strikingly good, 
given that no parameters are fitted to reproduce the experimental data. Not 
only are the principal peaks 2, 6 and 12 well reproduced, but the 
relatively large peaks at 4, 9, and 16 for the high-spin states at half-shell 
filling are also clear~\cite{koskinen,reimann1}. For $N=4$, Hund's first 
rule correctly predicts the calculated $S=1$ spin-triplet state in 
which spins are aligned in the two highest partially occupied degenerate 
single-particle levels ($n$,$l$)=(0,1) and (0,-1), rather than the $S=0$ 
spin-singlet state in which the paired-spin electrons reside in either 
the (0,1) or (0,-1) levels.

	Deforming the confinement slightly by changing the deformation 
parameter to $\delta =1.1$ (see trace (b) in Fig.~\ref{fig:4}), the 
calculation still predicts fairly clear shell closures at $N=2,6$ and 12. 
These numbers can still be considered as ``magic'', but the actual values of 
$\Delta _2(2,6,12)$ are noticeably suppressed, because degeneracies have
been lifted~\cite{TARUCHA2}. The $N=20$ peak has become very weak. 
Also values of $\Delta _2(N)$ neighboring $N= $2, 6, and 12 start to become 
comparable to the values for $N=2,6,$ and 12, i.e. there is less contrast. 
Overall, the shell structure is much less pronounced compared to that for 
the circle. Already it is clear that even a very small deviation from perfect 
circular symmetry can have a very noticeable effect even when single-particle 
level degeneracies are lifted by just a small amount.

	As the deformation increases further, the pronounced peaks for 
$N=$2, 6, 12, and 20 evident for the disk-shaped dot are further suppressed. 
This is a simple consequence of the removal of the level ``bunching'' with 
deformation. Even for the cases where ``accidental'' subshell closures occur 
at certain ``magic'' deformations (e.g. $\delta =1.5$ and 2 as seen in 
Fig.~\ref{fig:2}), the reduced separation between degenerate single-particle 
energy levels ($E_L$) would make any shell structure less clear to observe, 
and the sequence of ``magic'' numbers would be very different (e.g. for 
$\delta =2$ it would be 2, 4, 8, 12, 18, ...) compared to those for 
$\delta =1$. From Fig.~\ref{fig:4} we can see essentially that for 
$\delta \ge 1.2$, the circular shell structure has been completely 
eliminated. Traces (a) to (f) thus illustrate the dramatic destruction of the 
familiar shell structure for a circular dot with deformation.  

	Also apparent is that a systematic one-to-one correspondance of 
$\Delta _2(N)$ between traces (b) to (d) in Fig.~\ref{fig:3} and traces 
(b) to (f) in Fig.~\ref{fig:4} is impossible to make. Although the experimental 
data for mesa X partly resembles the theoretical data for $\delta $= 1.1 
to 1.3, the data for mesas Y and Z do not seem to resemble that for 
$\delta >1.3$, except perhaps for a weak tendency to oscillate between 
even-$N$ and odd-$N$. We have already stated many reasons why, in comparision 
to a circular dot, a good correspondance between experiment and theory for 
the elliptical dots is less likely. We stress that ultimately, except for 
circle W, $\delta $ is not known, and equating $\delta $ with 
$\beta $ may not be reliable. To progress we must look for other clues.

	Theoretically, Fig.~\ref{fig:4} shows that there are transitions in 
the ground state spin-configurations with deformation~\cite{reimann1}. The 
total spin, $S$, is identified by different symbols in the figure. These 
transitions are particularly numerous for, but are not restricted to, 
the even-$N$ systems, and are clearly very sensitive to the actual value of
the deformation. For example, in the case of $N=6$ electrons, the total spin 
is predicted to change from $S=0$ (i.e. a paramagnetic state) at $\delta =1$, 
through an $S=0$ SDW state, to $S=1$ at $\delta =1.5$ --- an indication of 
``piezo-magnetic'' behavior~\cite{reimann1,mss8},~i.e.~changes of the 
dot magnetization with deformation. Although experimentally we are not in a 
position to differentiate between an $S=0$ ``normal'' state and an $S=0$ 
SDW state showing a spatial variation in the polarization as a consequence 
of broken spin symmetry in the internal coordinates
~\cite{ringschuck}- indeed the interpretation of a SDW is still debated in 
the literature~\cite{Hirose}- the SDFT calculations described here predict 
that the latter becomes more prevalent for even-$N$ systems as $\delta $ 
increases, particularly for small average particle densities
~\cite{koskinen,reimann1}. 

	Another interesting, and in practice the simplest incidence we can 
focus on, is what happens to the $N=4$ ground state. The inset in 
Fig.~\ref{fig:4} shows $\Delta _2(N=4)$ versus deformation up to 
$\delta =1.5$. Starting with the circular dot, Hund's first rule gives a 
total spin of $S=1$ for the triplet state favoring spin alignment of the two 
electrons in the second shell rather than a total spin of $S=0$ for the 
singlet state in which the spins are paired. As the deformation is initially 
increased, the energy separation between the two levels ($n_L$,$n_S$)= (1,0) 
and (0,1) -- the two originally degenerate levels ($n$,$l$)= (0,1) and (0,-1) 
in the second shell of the circular dot -- increases (see (a) and (b) in
Fig.~\ref{fig:2}), and so the spin-triplet state becomes progressively less 
favorable. $\Delta _2(4)$ continously decreases with $\delta $, and at a value 
between 1.2 and 1.3, a spin-zero state (actually predicted by the SDFT 
described here to be a SDW) appears, i.e. a spin triplet-singlet transition 
is expected. For higher values of $\delta $ beyond this transition, 
$\Delta _2(N=4)$ starts to increase.

	Other recent calculations employing numerical diagonalization 
for elliptical dots moderately deformed up to $ \delta =2$ have also predicted 
that $\Delta _2(N)$ is sensitive to deformation, and that the spin-states can 
be modified~\cite{EZAKI,EZAKI2}. Those calculations, for $N$ up to 10, and 
performed at 0 T with $E_Q$=3 meV, also reveal a spin triplet-singlet 
transition at $\delta \approx 1.2$ for $N=4$, and, more generally, a 
consecutive filling of states by spin-up and spin-down electrons at higher 
deformation is favored.

	Inspection of Fig.~\ref{fig:3} gives values of $\Delta _2(N=4)$ for 
mesas W, X, Y, and Z respectively of 3.1, 2.7, 3.1, and 2.5 meV. Whilst it is
reassuring that these energies lie in the range predicted by SDFT, it is 
tempting to attribute, for a $\delta $ value equated to the $\beta $ 
value, the apparently anomalously low value for mesa Z to sample specific 
fluctuations, and say that the trend for mesas W, X, and Y is consistent 
with that predicted in Fig.~\ref{fig:4}~({\it inset}),~i.e.~$N=4$ is a spin-
triplet for W, and a spin-singlet for X, Y, and Z. However, as we do not 
really know $\delta$ for the elliptical dots, we can not even be confident 
that the actual $\delta$ values lie in the $\delta $=1.0 to 1.5 range, i.e. 
the values might be higher, or even that the order X, Y, and Z for increasing 
deformation as suggested by the $\beta $ values is correct. Fortunately, 
we can apply a $B$-field, and as we will shortly show this goes a long way 
to resolving these difficult issues.

	In case the actual $\delta$ values for the elliptical dots exceed 1.5,
traces (g) and (h) in Fig.~\ref{fig:4} respectively show $\Delta _2(N)$ for 
the higher deformation parameters $\delta =2$ and $\delta =3.2$. We have no 
reason to believe that $\delta $ experimentally will be exactly 2 or exactly 
3.2, but the numbers are representative of the two situations where, 
respectively, many or no single-particle levels are degenerate at 0 T for 
non-interacting electrons, as illustrated by the spectra in Fig.~\ref{fig:2}. 
As expected, traces (g) and (h) show no circular-like shell structure, and no 
particularly large values of $\Delta _2(N)$. Indeed, apart from the 
``classical'' background trend, i.e. $\Delta _2(N)$ increasing as $N$ 
decreases, there is little one can say about the traces except for $N>5$ 
there is a tendency for a weak even-odd oscillation in $\Delta _2(N)$, and 
this oscillation is perhaps clearer for larger $\delta $. The model here 
actually predicts small peaks for odd-$N$, and small valleys for even-$N$. 
For odd-$N$ the spin-state is nearly always S=1/2, and for even-$N$ the 
spin-state is usually $S=0$ (SDW). At least for $\delta =2$, where 
in the single-particle picture there can be accidental degeneracies 
at 0 T (see Fig.~\ref{fig:2} trace (c)), one might naively expect 
some non-zero even-$N$ spin-states, but it is possible that in the model 
calculations, for the parameters given, the interactions modify the spectrum 
so dramatically that expected degeneracies are lifted reducing the visibility 
of any potential shell structure, e.g.~$N=6$ and 10 are predicted here to be
$S=0$ (SDW) rather than $S=1$ as might be expected from Hund's first rule. 
On the other hand, for $\delta =3.2$, where in the single-particle picture 
there are no accidental degeneracies at 0 T (see Fig.~\ref{fig:2} trace (d)), 
perhaps surprisingly some non-zero even-$N$ spin-states, for example for 
$N$=12 and 16, are predicted -- this too may be due to interactions. 

	For Y and Z, the $\Delta _2(N)$ traces in Fig.~\ref{fig:3}  
seem to show a weak tendency to oscillate between a slightly larger 
even-$N$ value, and a slightly smaller odd-$N$ value, and this oscillation 
seems clearer for Y than for Z. For the moment we do not try to account 
for the clarity of this oscillation in dots Y and Z, but try to explain the 
origin of the oscillation, although we are now being forced to entertain the 
idea that $\delta $ for Y and Z may be much larger than 1.5. Starting from 
the over simple single-particle picture with a fixed confinement energy, and 
then including a constant interaction which is the same for even-$N$ and 
odd-$N$, a larger even-$N$ value is expected because only 
$\Delta _2($even-$N)$ can contain a finite contribution due to the 
single-particle energy level spacing. A slightly more advanced model, 
which is more realistic in principle, would be to have a constant 
interaction for odd-$N$ (next electron added to an $S=1/2$ state already 
containing one electron) that is stronger than the constant interaction 
for even-$N$ (next electron added to an empty state). 
If the former is larger than 
the latter plus the single-particle spacing (more likely in practice as 
$N$ increases), a weak tendency to oscillate between smaller even-$N$ and 
larger odd-$N$ could occur. This pattern is what the SDFT calculations 
predict in Fig. 4 for $\delta =2 $ and $\delta = 3.2$. The fact that 
$\Delta _2(N)$ for Y and Z is often a little larger for even-$N$ than 
odd-$N$ should not be taken to mean that the constant interaction model is 
more accurate. Rather the Coulomb interactions may not be so strong in 
practice, due to screening by the leads for example, as those in our 
model- a model that also does not include the self-consistent calculation 
of the electrostatic confining potential. Indeed, in the SDFT 
calculations of Lee et al.~\cite{LEE}, the electrostatic confining potential 
is much stronger (e.g. $E_S$=20 meV, $E_L$=10 meV), and they find that 
$\Delta _2(N)$ is generally a little larger for even-$N$ than for 
odd-$N$. Finally, we note that eventually, for a much stronger deformation 
(e.g. $\delta $ exceeding 10), the addition energy spectrum would become 
smoother as it tends towards that for a quasi-one-dimensional quantum 
wire~\cite{reimann1,reimann2}. 

\section{Magnetic Field Dependence}

	Application of a magnetic field is a powerful tool with which to 
identify the quantum numbers of states in our vertical quantum dots
~\cite{TARUCHA2,TARUCHA,KOUWENHOVEN}. Fig.~\ref{fig:2} instructively shows 
the expected evolution of the first ten single-particle energy levels 
with $B$-field up to 6 T for a circular dot ($\delta $=1), and for 
elliptical dots with $\delta=1.5, 2,$ and 3.2. The energy level spectra are 
calculated according to the simple single-particle model employed by 
Madhav and Chakraborty\cite{MADHAV} in which Coulomb interactions are 
neglected, and the confining potential is assumed to be perfectly parabolic. 
The spectrum for the circular dot is the familiar Darwin-Fock 
spectrum for a circular two-dimensional harmonic confining potential. The 
confinement energy for the circular dot, $E_Q$, is taken to be 3 meV, 
in practice a reasonable average value in the few-electron limit, and is 
assumed to be independent of $N$. The confinement energies for the 
elliptical dots are simply derived from the relation $E_L$$E_S$=$E_Q$$E_Q$. 
For the case of the circle and the $\delta=3.2$ ellipse, quantum numbers 
($n$,$l$) and ($n_L$,$n_S$) respectively for some of the states we discuss 
are indicated. Each single-particle energy level can accomodate a spin-up 
and spin-down electron, so current peaks should normally come in pairs in 
a constant interaction model neglecting exchange~\cite{MADHAV}. ``Wiggles''
in the position of pairs of current peaks are expected because the 
$B$-field induces crossings between single-particle states
~\cite{TARUCHA2,TARUCHA,MADHAV}. The first lowest energy ``wiggle'' 
originates from the crossing marked by a black triangle in each of the four 
spectra. $\Delta _2$(even-$N$) is expected to be strongly dependent on $B$-
field as it can contain contributions from single-particle energy level 
spacings, whereas $\Delta _2$(odd-$N$) is essentially independent of 
$B$-field at weak-field, and is determined only by the effect of Coulomb 
repulsion. Any detailed discussion on the actual $B$-field dependence of the 
current peaks requires the inclusion of Coulomb interactions
~\cite{KOUWENHOVEN}. The four calculated spectra nevertheless clearly serve 
to demonstate three simple points: $i)$ the $B$-field lifts all degeneracies 
present at 0 T at the ``magic'' deformations, e.g. $\delta =$1, 1.5, 2, ... 
($\delta$=3.2 is not a ``magic'' deformation); $ii)$ a $B$-field can always
induce degeneracies at finite field when single-particle levels cross,
provided the confinement potential is perfectly parabolic; and $iii)$ as 
$\delta $ increases, the single-particle energy level spacing generally 
decreases ($\le E_L$).

	Fig.~\ref{fig:5} shows the $B$-field dependence, for a weak field 
applied parallel to the current, of the Coulomb oscillation peak positions for 
the circular mesa W, (a), and the rectangular mesas X, Y and Z, (b) to (d). 
The data consists of current vs. $V_g$ traces taken at a very small bias 
($\ll 1$~mV) at different $B$-fields at or below 0.3~K. 

	For circle W, only the third, fourth, fifth and sixth current peaks 
(belonging to the second shell at 0 T) are shown. The pairing of the third 
peak with the fifth peak, and the fourth peak with the sixth peak from 0 T 
to 0.4 T, as opposed to the more usual pairing of the third peak with the 
fourth, and the fifth peak with the sixth (due to consecutive filling of
electrons into spin-degenerate single-particle states) for B$>$0.4~T, is a
consequence of Hund's first rule: the $N=4$ state is a spin-triplet so two 
parallel-spin electrons fill the two different but originally degenerate 
states ($n$,$l$)=(0,1) and (0,-1) in the half-filled second shell
~\cite{TARUCHA2,TARUCHA}. For B$>$0.4~T, the fifth and sixth peaks, as a pair, 
first move up, as indicated by the thick arrow, and then start to move down 
at about 1.4 T due to the crossing of the single-particle states ($n$,$l$)=
(0,-1) and (0,2). This lowest single-particle level crossing, which is also 
clear in Fig.~\ref{fig:2}(a), is marked by a black triangle. The spins of 
the added electrons are also shown pictorially at 0 T and 2 T. 

       To explain why Hund's first rule is obeyed in a simple way, 
we can introduce an energy, $E_{EX}$, to represent the reduction in energy 
due to exchange between electrons in the half-filled second shell, and this is
estimated to be about 0.7 meV for circle W~\cite{TARUCHA2,TARUCHA}. The $N=4$ 
triplet-state is thus lower in energy than the $N=4$ singlet-singlet state 
by $E_{EX}$, and as a consequence $\Delta _2(3),\Delta _2(5)<\Delta _2(4)$ by 
about 2$E_{EX}$. This exchange-related effect persists in a weak B-field as 
long as the splitting between states $(0,1)$ and $(0,-1)$ is less than 
$E_{EX}$. At 1.4 T this splitting exceeds $E_{EX}$, and the ground state 
becomes a spin-singlet, i.e. there is a B-field induced triplet-singlet 
transition.

	For rectangles X, Y, and Z, the first ten current peaks are shown 
in Fig.~\ref{fig:5},(b) to (d). Peaks are paired, and there are no obvious 
deviations close to 0 T for $N=4$ which can be attributed to exchange effects,
i.e. Hund's first rule. Quantum numbers ($n_L$,$n_S$) of the single-particle 
states are assigned, and the first up-moving pair of peaks is marked by a 
thick arrow. With increasing deformation, the first up-moving pair of 
peaks, and the lowest energy single-particle level crossing- identified by a 
black triangle in each of the Fig.~\ref{fig:2} spectra- are simply expected to 
move systematically to higher $N$ (or equivalently to higher energy)
~\cite{MADHAV}. 

	For the elliptical dots, normal peak pairing, even from 0 T, occurs 
so Hund's first rule is not obeyed. This suggests that the spin-state 
for $N=4$ is a singlet. The exchange effect is maximal for a circular dot at 
$N=4$ because the ($n$,$l$)=(0,1) and (0,-1) states are degenerate, but with 
deformation these states become the ($n_L$,$n_S$)=(1,0) and (0,1) states in 
an elliptical dot which are split at 0 T. This energy splitting, $\gamma $, 
increases with $\delta$. If $\gamma < E_{EX}$ at 0 T, exchange can still 
operate to lower the energy, and thus the $N=4$ ground state remains a 
spin-triplet. On the other hand, if $\gamma > E_{EX}$ at 0 T, the energy gain 
due to exchange is not sufficiently large to compensate for the spliting,
so normal pairing occurs. Thus, as $\delta$ increases, we can expect a 
triplet-singlet transition at some critical deformation~\cite{SASAKI}. 
Note that $E_{EX}$ itself decreases with increasing deformation, as it has 
its maximum value only when the orbitals involved 
have the same symmetry. 
This transition is clear in the inset of Fig.~\ref{fig:4} according to SDFT, 
and has also been predicted by exact numerical diagonalization
~\cite{EZAKI,EZAKI2}. The tell-tail pattern in the trend of $\Delta _2(N=4)$ 
at 0 T with deformation should be an initial decrease while the state 
remains a spin-triplet, a turning point at the transition, and a rise 
thereafter when the state is a spin-singlet. As noted before, it is hard to 
judge from the absolute values of $\Delta_2(4)$ at 0 T alone shown in 
Fig.~\ref{fig:3} whether the $N=4$ state is a triplet or singlet. 
$\Delta_2(4)$ can be relatively large either side of the turning point if 
either $E_{EX}$ or $\gamma $ is large, i.e. a large $\Delta_2(4)$ can mean 
Hund's first rule is operating for nearly degenerate states, or there is a 
large separation between non-degenerate states. This potential ambiguity is 
apparent when we see that $\Delta_2(4)$ for circle W and ellipse Y are 
essentially equal, so it is vitally important to examine carefully the $B$-
field dependence. The absense of deviations to the normal peak pairing at 
$N=4$ in Fig.~\ref{fig:5}, traces (b) to (d), nevertheless does apparently 
confirm that $\delta $ is indeed greater than 1.2-1.3 
which is in line with the 
$\beta $ values for mesas X, Y, and Z. For completeness, we note that 
normally we probe the spin-states in our high symmetry dot structures via 
the orbital effect with the $B$-field parallel to the current. The spin-states
in the elliptical dot X have also been confirmed directly by measuring the 
Zeeman effect alone by applying a $B$-field perpendicular to the current, and 
the results are again consistent with a spin-singlet interpretation for $N=4$
~\cite{SASAKI}.

	The next most striking feature about traces (b) to (d) in 
Fig.~\ref{fig:5} is the position of the first up-moving pair of peaks. For 
mesas X, Y, and Z respectively, it is the 3rd, 5th, and 4th pair of peaks.
As revealed by the sequence of spectra in Fig.~\ref{fig:2}, in a simple 
single-particle picture, the first up-moving state is ($n_L$,$n_S$)=(0,1), 
which is actually the lowest energy state of the 
second Landau-level~\cite{MADHAV}. 
Inspection of these calculated spectra shows that this state is, in the weak-
field limit, from the bottom, the 3rd, 4th, and 5th state respectively for 
$1\le\delta <2$, $2\le \delta <3$, $3\le\delta<4$. Thus, starting from no
deformation, the first up-moving pair of peaks should go from the 3rd to 
4th, 4th to the 5th, ... at certain ``magic'' deformations as $\delta $ is
increased. Remembering that Coulomb effects are neglected in this simple 
picture, and that in practice $\delta $ is expected to vary with $N$, 
nonetheless, with these simple arguements it looks as if $1<\delta <2$ for X, 
$3<\delta <4$ for Y, $2<\delta <3$ for Z. If we believe this, then even 
though ellipses X, Y, and Z are all deformed beyond the triplet-singlet 
transition, we are forced to conclude the following: $i)$ $\delta $ can be 
much higher than that suggested by the $\beta $ values (especially for Y and 
Z); and $ii)$ the ordering given by increasing $\beta $ values may not
reflect the true ordering in $\delta $, i.e. the deformation in Y seems to be
stronger than in Z, so the true sequence may be W--X--Z--Y for the four mesas
considered. Given our earlier comments, the former is not so unexpected 
since we have no independent way of measuring $\delta $, but the latter is 
perhaps more surprising.

	If the true ordering of the ellipses is X, Z, and Y, even though the
reason for the deformation in Y being stronger than in Z is unclear to us, 
at least other attributes of Y and Z, and trends in Fig.~\ref{fig:3} and 
Fig.~\ref{fig:5} are consistent with this interpretation. For instance, 
$\Delta _2(N=4)$ for a W-X-Z-Y ordering respectively of 3.1, 2.7, 2.5, and 
3.1 meV is more in line with the predicted trend shown in the inset in 
Fig.~\ref{fig:4}, although the value for Z still seems a little low. This 
reordering does not contradict our earlier conclusion that, with the absense 
of deviations to normal pairing, $N=4$ is a spin-singlet state for all
three ellipses. Thus $\Delta_2(4)$ increases after the triplet-singlet 
transition, because with deformation the degeneracy of single-particle states 
is strongly lifted. The reversed ordering of 
ellipses Y and Z, as well as higher $\delta $ values than suggested by the 
$\beta $ values, also fits with the observations made earlier about 
traces (c) and (d) in Fig.~\ref{fig:3}, and traces (g) and (h) in 
Fig.~\ref{fig:4}. Both mesas show a tendency for $\Delta _2(N)$ to oscillate 
between slightly higher and slightly lower values respectively for even-$N$ 
and odd-$N$ and this seems more pronounced for Y than Z.

\section{Conclusions}

	We have experimentally and theoretically investigated the effect of 
ellipsoidal deformation on the shell structure, addition energies, and 
spin-states in vertical quantum dot atoms on going from circular- to 
rectangular-shaped mesas. The familiar and distinctive shell structure 
as determined from the addition energy spectra at 0 T for the circular dot is 
absent in the elliptical dots, and even small deviations breaking circular 
symmetry have a dramatic effect. Measurements with a magnetic field applied
parallel to current confirm that the $N=4$ spin state at 0 T has undergone a 
transition due to the moderate deformation: for the circular dot it is a 
spin-triplet in accordance with Hund's first rule when the second shell is 
half-filled, and for the elliptical dots it is a spin-singlet. These 
observations are in agreement with recent theory, as well demonstrated here 
by the application of spin-density functional theory at 0 T with a wide 
range of deformation parameters. The $B$-field dependence strongly suggests 
that the anisotropy of an elliptical dot in practice can be significantly 
higher than that given by simply considering the geometry of the mesa in 
which the dot is situated. In the future it will be interesting to experiment 
with even more strongly- and extremely-deformed dots (to clarify for 
instance the existance of $S=0$ SDW states), quasi-one-dimensional wire-like 
dots~\cite{reimann1,reimann2}, and possibly other exotically-shaped dots, 
for example ring-shaped dots~\cite{reimann2}, and triangular-shaped 
dots~\cite{EZAKI,EZAKI2}. Finally, with these goals in mind, ultimately 
better control and in-situ manipulation of the lateral potential geometry of 
a quantum dot is highly desirable, and this may be achieved by fully 
exploiting a multiple-gated vertical single electron transistor we have 
recently developed~\cite{AUSTING3,AUSTING5}.

\acknowledgments

	We would like to acknowledge the considerable assistance of Takashi 
Honda in the fabrication of the devices, and useful discussions with 
Yasuhiro Tokura, and Hiroyuki Tamura. This work is partly supported by 
grant 08247101 from the Ministry of Education, Science, Culture and Sports, 
Japan, the NEDO joint research program (NTDP-98), the Academy of Finland, 
and the TMR program of the European Community under contract 
ERBFMBICT972405. Our understanding of the circular dots has benefitted from
a long term collaboration with Leo Kouwenhoven at the Delft University of
Technology and his coworkers.

\begin{figure}
\caption{Typical scanning electron micrographs show a circular mesa with top 
contact diameter D, and a rectangular mesa with top contact area 
$L\times S \quad(L>S)$ taken immediately after the deposition of the Schottky 
gate metal surrounding the mesa. The slight undercut due to the light wet 
etch is clearly visible. Schematic diagrams depict the slabs of semiconductor 
between the two tunneling barriers, and the resulting circular and elliptical
shaped dots bounded by the shaded depletion region for the circular and 
rectangular mesas respectively. The current flows vertically through the dots 
in the direction indicated. For the rectangular mesa, the energy parabola 
along the major and minor axes are included and the respective confinement 
energies, $E_L$ and $E_S (E_L<E_S)$, are marked.}
\label{fig:1}
\end{figure}
\begin{figure}
\caption{Magnetic ($B$-) field dependence up to 6 T of the first ten single-
particle energy levels (each level can hold a spin-up and spin-down
electron) for a circular dot with $\delta =1$, (a), and for elliptical dots 
with $\delta= $1.5, 2, and 3.2, (b) to (d). The energy level spectra are 
calculated with the simple formalism employed by Madhav and Chakraborty, 
as explained in the text. 
The spectrum for the circular dot is in fact the familiar 
Darwin-Fock spectrum for a circular two-dimensional perfectly harmonic 
confining potential. The confinement energy for the circular dot, $E_Q$, is 
taken to be 3~meV, a reasonable average value in the few-electron limit, 
and is assumed to be independent of $N$. The confinement energies for the 
elliptical dots are simply derived from the relation $E_LE_S=E_QE_Q$. For 
the circular dot and the $\delta =3.2$ ellipse, quantum numbers ($n,l$) 
and ($n_L,n_S$) respectively, for some states are given. At 0 T accidental 
degeneracies are also evident for the $\delta =1.5$ and $\delta =2$ ellipses, 
but not for the $\delta =3.2$ ellipse. The black triangles mark the position 
of the first (lowest energy) single-particle crossings}
\label{fig:2}
\end{figure}
\begin{figure}
\caption{Change in the electro-chemical potential, $\Delta _2(N)$, as a 
function of electron number, $N$, up to 17 at 0 T for circular mesa W of 
diameter 0.5 $\mu $, (a), and three rectangular mesas X, Y, and Z of area 
0.55 x 0.4  $\mu$m$^2$, 0.65 x 0.45  $\mu$m$^2$, and 0.6 x 0.4  $\mu$m$^2$ 
respectively, (b) to (d). For W, $\delta=1$, and for X, Y, and Z 
respectively, $\beta $ is nominally 1.375, 1.44, and 1.5. The traces are 
offset vertically by 3~meV for clarity. For circular dot W, a clear shell 
structure is observed. Peaks due to full ($N=2$,6,12) and half-full 
($N=4$,9,16) shell filling are numbered, and the fit given by local spin-
density approximation (LSDA) is included, as discussed in the text and also 
shown in Fig.~\ref{fig:4}}
\label{fig:3}
\end{figure}
\begin{figure}
\caption{Model calculations for the change in electro-chemical potential, 
$\Delta _2(N)$ within spin-density functional theory. The different traces 
correspond to zero, weak and moderate deformation parameters $\delta =1.0$ 
to $\delta =1.5$, (a) to (f), and higher deformation parameters $\delta =2$ 
and $\delta =3.2$, (g) and (h). The traces are offset vertically by 1~meV for 
clarity, and there is an additinal 1~meV offset between traces (f) and (g). 
Traces (a) to (f) illustrate well the dramatic destruction with deformation 
of the familiar shell structure for a circular dot. The total spin, $S$, for 
different deformations and electron numbers are identified by different 
symbols as defined in the figure. Note that with increasing deformation, 
$S=0$ spin-density wave (SDW) states are predicted to become more prevalent 
for even-$N$. The inset shows $\Delta _2(N=4)$ versus $\delta$. For 
$\delta<1.2$ and $\delta>1.3$ respectively the $N=4$ ground state is expected 
to be a spin-triplet (S=1) and spin-singlet $(S=0)$.}
\label{fig:4}
\end{figure}
\begin{figure}
\caption{Magnetic ($B$-) field dependence of the Coulomb oscillation peak 
positions for mesas W, X, Y and Z, (a) to (d). The $B$-field is parallel to 
the current. The data consists of current versus $V_g$ traces for different 
$B$-fields which are offset, and rotated by 90~degrees. A less negative 
gate voltage corresponds to higher energy. For circular dot W, only the 3rd, 
4th, 5th, and 6th current peaks belonging to the second shell are shown. The 
unusual pairing of the 3rd peak with the 5th, and the 4th peak with the 6th 
from 0 to 0.4 T as opposed to the more usual pairing for B$>$0.4~T 
is evident, and is related to Hund's first rule. For B$>$0.4 T, the 5th and 
6th peaks, as a pair, first move up, as indicated by the thick arrow, and 
then start to move down at about 1.4 T due to the crossing between the 
single-particle states ($n$,$l$)=(0,-1) and (0,2). The spins of the added 
electrons are shown pictorially at 0 and 2 T. For elliptical dots X, Y, Z, 
the first ten current peaks are shown. Peaks are paired, and there are no 
obvious deviations close to 0 T for $N=4$ which can be attributed to 
exchange effects. Quantum numbers ($n_L$,$n_S$) of the single-particle 
states are included, and the first up-moving pair of peaks is marked by a 
thick arrow. As deformation increases, the single-particle spectra in 
Fig. 2 show that this up-moving pair should move to higher $N$.}
\label{fig:5}
\end{figure}
\end{multicols}

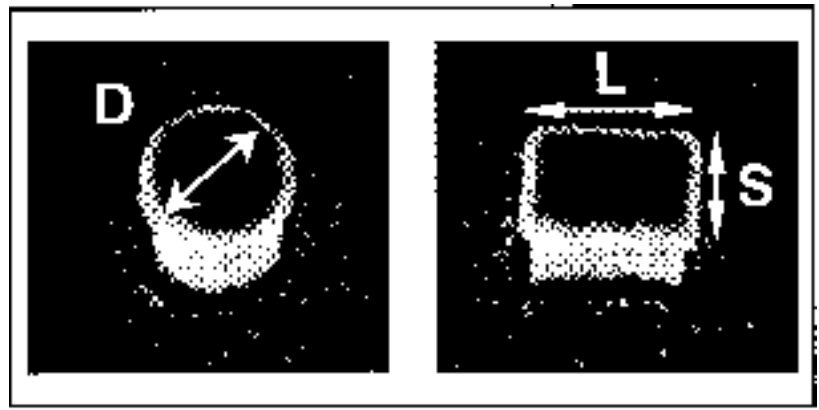
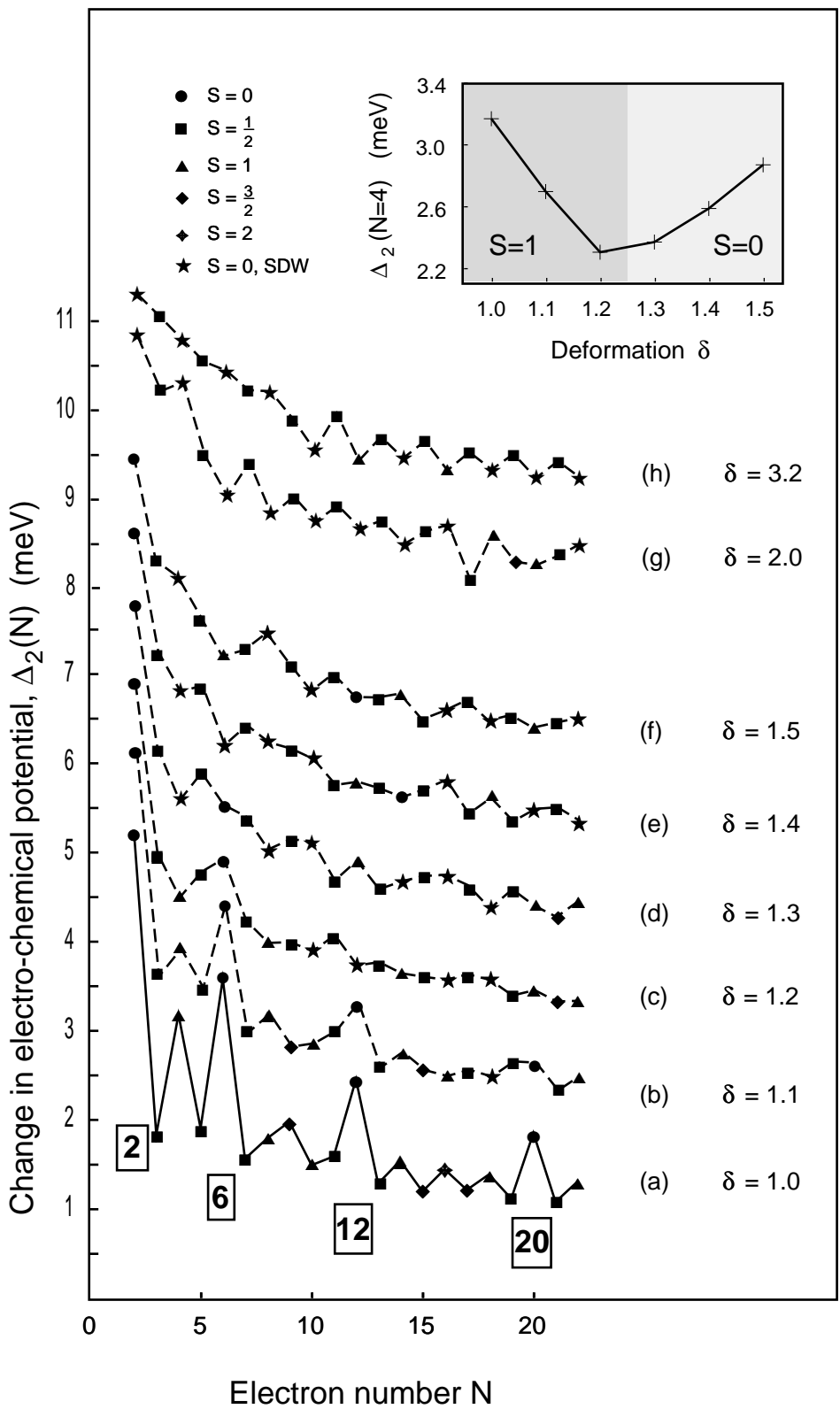
\end{document}